\documentstyle[twoside,fleqn,npb,epsfig]{article}
\newcommand{\bea}{\begin{eqnarray}}
\newcommand{\eea}{\end{eqnarray}}
\newcommand{\non}{\nonumber}

\newcommand{\AmS}{{\protect\the\textfont2
  A\kern-.1667em\lower.5ex\hbox{M}\kern-.125emS}}

\title{Higher order QED corrections to deep inelastic scattering}

\author{J. Bl\"umlein and H. Kawamura\address{Deutsches
        Elektronen Synchrotron, DESY, Platanenallee 6, D--15738 
        Zeuthen, Germany}\thanks{Supported in part by EU--TMR Network
        HPRN-CT-2000-00149}}

\begin{document}

\begin{abstract}
\noindent
We calculate the leptonic $O(\alpha^2 L)$ QED corrections for unpolarized
deeply inelastic $ep$ scattering using mixed variables.
\end{abstract}

\maketitle

\section{INTRODUCTION}

\noindent
Deep inelastic electron--nucleon scattering allows for fundamental QCD
tests investigating the scaling violations of structure functions
in the perturbative regime of large values of $Q^2$. The detailed 
knowledge of the structure functions enables to study various aspects of 
the dynamics of non-Abelian gauge theory, and is necessary for the future
experimental search for the Higgs--boson  and new particles at TEVATRON 
and LHC. 

One of the major goals of the experiments H1 and ZEUS at the 
$ep$--collider HERA at DESY is to perform a QCD test at large space--like
virtualities $Q^2$ at high precision. This presumes to know 
the QED radiative corrections to the double--differential scattering 
cross 
sections of deeply inelastic $ep$ corrections as precisely as possible.
Previous calculations of the radiative corrections for the unpolarized 
cross sections at leading order [1--5]\footnote{For QED corrections
to polarized lepton scattering off polarized nucleons see~\cite{POLA}.},
the leading--log level [7--12] to leading and higher orders, and
QED--resummations of small--$x$ terms~[11,13] revealed that these 
corrections are very large, in a wide kinematic range of $x$ and $Q^2$.
The corrections are, moreover, complicated
by a new type of sizeable contributions as the Compton--peak~\cite{COMPT}.
This makes it necessary to extend the calculations to higher orders.

The higher order leading--logarithmic contributions $O\left[(\alpha L)^k
\right]$ to QED corrections are obtained as the leading order solution
of the associated renormalization group equations~\cite{CS} for mass
factorization. These corrections are universal, process--independent
w.r.t. their structure, and are given in terms of Mellin--convolutions
of leading order QED splitting functions. The next--to--leading order 
(NLO) corrections can be obtained along the same line. However, besides 
the splitting functions to NLO also the respective process--dependent
Wilson coefficients and operator matrix elements in the on-mass-shell
(OMS) scheme contribute. In the past this method was applied to 
calculate the $O(\alpha^2)$ initial--state QED radiative corrections to 
$e^+e^- \rightarrow \mu^+\mu^-$ in 
Ref.~\cite{BBN}.

In the present paper we summarize results of a recent calculation
\cite{JBHK2}
of the leptonic QED corrections to deeply inelastic $ep$ scattering 
defining the double--differential scattering cross section for mixed 
variables~[8,2,5] to $O(\alpha^2 L)$. If compared to the case dealt with in \cite{BBN} to
$O(\alpha^2 L)$ the present calculation is more complicated due to the
emergence of final state radiation, the double--differential cross section
and the relevant rescaling, which    implies  non--Mellin type
convolutions in general. We first summarize main kinematic aspects and
present then the different contributions to the leptonic NLO QED 
corrections. 
\section{Mixed variables}

\noindent
$ep$ collider experiments allow to measure the kinematic variables
defining the inclusive deep--inelastic scattering cross sections in 
various ways since in principle four kinematic variables are available
with the energies and angles of both the outgoing lepton and the 
struck--quark, see e.g.~\cite{KIN}. At the Born level all methods are
equivalent, however, resolution effects as a consequence of the
detector's structure, differ in certain kinematic regions. The 
Bremsstrahlung--effects of QED radiative corrections change this picture 
drastically and the QED correction factors depend on the way the kinematic
variables, as e.g. Bjorken--$y$ and the virtuality $Q^2$ are 
measured.\footnote{See e.g. Refs.~[8,2,5] for a comparison a wide
range of different choices of measurement.} In the present paper
we calculate the NLO radiative corrections in the case of neutral
current deep--inelastic scattering for mixed variables, i.e. that 
$Q^2 = Q^2_l$ is measured at the leptonic and $y=y_h$ is measured at the 
hadronic vertex, and $x_m =Q^2_l/(S y_h)$.  The Born cross
section for $\gamma$--exchange is given by~:
\bea
\frac{d^2 \sigma^{(0)}}{dy dQ^2}
= \frac{2\pi \alpha^2}{y Q^4} \left[
y^2~2xF_1 + 2(1-y)~F_2\right],
\eea
with
\begin{eqnarray}
F_1(x,Q^2) &=& \frac{1}{2}
\sum_{k=1}^{N_f} \left[q_k(x,Q^2) + \overline{q}_k(x,Q^2)
\right],\\
F_2(x,Q^2) &=& 2x F_1(x,Q^2) + F_L(x,Q^2)~.
\end{eqnarray}
Here, $F_{1,2,L}(x,Q^2)$ denote the nucleon structure functions for
photon exchange, and $q(x,Q^2)$ and $\overline{q}(x,Q^2)$ are
the quark-- and antiquark distribution functions. The sub--system 
variables obey the following rescaling relations for initial-- and final--state 
radiation~:
\begin{eqnarray}
\label{eqRESC1}
&&\hspace{-0.6cm}{\sf ISR~:}~~~\widehat{y} = \frac{y_h}{z},~\widehat{Q}^2 
           = z Q^2_l,~\widehat{S} = zS,~\widehat{x} = zx_m,
           \nonumber \\
&&~~~~~J^I(z) = 1,~~~~~z_0^I = {\rm min}\left\{y_h,\frac{Q_0^2}{Q_l^2}
\right\}~,  \\
\label{eqRESC2}
&&\hspace{-0.6cm}{\sf FSR~:}~~~\widehat{y} = y_h,~\widehat{Q}^2
           = \frac{ Q^2_l}{z},~\widehat{S} 
           = S,~\widehat{x} = \frac{x_m}{z},
           \nonumber \\
& &~~~~~J^F(z) = \frac{1}{z},~~~~~z_0^I = x_m~.
\end{eqnarray}
Here, $J^{I,F}(z)$ are the initial-- and final--state Jacobians
$d^2(\widehat{y},\widehat{Q}^2)/d^2(y_h,Q^2_l)$, and $z_0$ marks the
lower bound of the sub--system rescaling variable $z~\epsilon~[z_0,1]$.
The rescaling in Eqs.~(\ref{eqRESC1},\ref{eqRESC2}) was chosen such, that
both the initial-- and final--state operator matrix elements can be
expressed with a variable $z~\epsilon~[0,1]$. $Q_0^2$ is introduced as
a scale to cut away contributions of the Compton peak. Although these
terms do formally belong to the QED radiative corrections, they stem
from a kinematic domain of low virtualities and are therefore not being
associated to deep inelastic scattering. The scale $Q_0^2$ can be chosen
by experiment accepting only those events in the sample to be analyzed
for which the {\it hadronic} $Q^2$ is larger than $Q_0^2$. By this measure
the Compton peak is cut away widely and the QED--correction factor is
dominated by a deep--inelastic sub--process by far. 
In the case of mixed variables the leptonic QED radiative corrections can
be easily grouped into those for the initial and final state. 
The separation scale between the two kinematic regions is $Q_l^2$.
\section{NLO corrections}

\noindent
In this paper we limit the consideration to the calculation of the
NLO corrections to leptonic variables for one--photon exchange in
electron--nucleon scattering. This approach is widely model independent
and allows to refer to general non--perturbative parameterizations of
the structure functions which describe the hadronic tensor. In this way
a direct unfolding of the experimentally measured structure functions is
possible down to the range in $Q^2$ and $x$ in
which partonic approaches fail 
to provide a description of structure functions.
The radiative corrections calculated are thus valid as well for inclusive
{\it diffractive} $ep$--scattering, see e.g.~\cite{DIFFR}. Here the
$k$-th order cross section is denoted by,
\begin{eqnarray}
\label{eq2L}
\frac{d^2 \sigma^{(k)}}{dy_h dQ_l^2} \hspace{-1mm} = \hspace{-1mm}
\sum^k_{l=0}\left(\frac{\alpha}{2\pi}\right)^k \hspace{-1mm}
\ln^{k-l} \hspace{-1mm}
\left(\frac{Q^2}{m_e^2}\right)C^{(k,l)}(y,Q^2)
\non\\
\end{eqnarray}
with $C^{(0,0}(y_h,Q_l^2) ={d^2 \sigma^0}/{dy_h dQ_l^2}$.
The  $O\left[(\alpha L)\right]$ and $O\left[(\alpha L)^2\right]$ 
corrections were calculated in Ref.~\cite{JB94}. The term 
$C^{(1,1)}(y_h,Q_l^2)$ was derived in Ref.~\cite{ABKR} 
completing the $O(\alpha)$ corrections. We re-calculated 
these corrections and agree with the previous results.

The NLO--correction $C^{(2,1)}(y_h,Q_l^2)$ can be obtained representing 
the scattering cross section using mass--factorization. Although the 
differential scattering cross section does not contain any mass 
singularity, one may decompose it in terms of Wilson coefficients and
operator-matrix elements being convoluted with the Born cross section.
In this decomposition both the operator matrix elements and the Wilson
coefficients depend on the factorization scale $\mu^2$. 
One writes the scattering
cross section as, see also \cite{BBN}\footnote{Very recently also 
the electron
energy spectrum
in muon decay was calculated~\cite{AM} using this method.},
\begin{eqnarray}
\frac{d^2 \sigma}{dy_h dQ_l^2}=
\frac{d^2 \sigma^0}{dy_h dQ_l^2} 
\otimes 
\sum_{i,j}\Gamma^I_{ei} \otimes \hat{\sigma}_{ij} \otimes \Gamma^F_{je}
\end{eqnarray}
with $\Gamma^{I,F}_{ij}(z,\mu^2/m^2_e)$ the initial and final state
operator matrix elements and $\hat{\sigma}_{kl}(z,Q^2/\mu^2)$ the 
respective
Wilson coefficients. $\otimes$ denotes a convolution, which depends
on specific rescalings of the chosen kinematic variables for the
differential cross sections. Both the
operator matrix elements and the Wilson coefficients obey the
representations
\begin{eqnarray}
&&\hspace{-0.7cm}
\Gamma^{I,F}_{ij}=
\delta(1-z) + \sum_{m\geq n} \hat{a}^m\Gamma_{ij}^{I,F(m,n)}L^{(m-n)} \\
&&\hspace{-0.5cm}
\hat{\sigma}_{kl}=
\delta(1-z) + \sum_{m\geq n} \hat{a}^m \widehat{\sigma}_{kl}^{(m,n)}
\widetilde{L}^{(m-n)},
\end{eqnarray}
where $\hat{a}=\alpha/(2\pi)$ and the sequences $\{ij\}$ and $\{kl\}$
in the above
do always denote $j(l)$ for the incoming and $i(k)$ the outgoing
particle, and $L$, $\widetilde{L}$ denote 
$\ln\left({\mu^2}/{m^2}\right)$, $\ln\left({Q^2}/{\mu^2}\right)$
respectively.
As the differential cross section is 
$\mu$--independent, the cross section is expressed by  convolutions
of the functions $\Gamma_{ij}^{I,F(m,n)}(z)$ and  
$\widehat{\sigma}_{kl}^{(m,n)}(z)$ such,
that the $\mu^2$--dependence cancels and a structure like
in Eq.~(\ref{eq2L}) is obtained. The present treatment in the OMS scheme
assumes that the light fermion mass, $m_e$, is kept everywhere it is 
giving a final answer in the scattering cross section if compared to the 
large scale $Q^2$, i.e. the only terms being neglected are power 
corrections which are of $O\left[(m_e^2/Q^2)^k\right],~k \geq 1$ and 
therefore small. The last step is necessary to maintain the anticipated 
convolution structure which, in parts, is of the Mellin--type, as also in
a massless approach.

In the subsequent relations we make frequent use of the rescaling 
(\ref{eqRESC1},\ref{eqRESC2}). For this purpose we introduce the
following short--hand notation for a rescaling a function $F(y,Q^2)$
\begin{eqnarray}
\label{RESCA}
\widetilde{F}_{I,F}(y,Q^2) 
= F\left(y=\widehat{y}_{I,F},Q^2=\widehat{Q}^2_{I,F}\right)~,
\end{eqnarray}
where $I,F$ label the respective type of rescaling.
The NLO--corrections may be grouped into the following contributions~:
 
\vspace{2mm}\noindent
{\sf
\renewcommand{\labelenumi}{\theenumi}
\begin{enumerate}
\item[i~] LO initial and  final state radiation off 
$C^{(1,1)}_{ee}(y,Q^2)$
\item[ii~]coupling constant renormalization of $C^{(1,1)}_{ee}(y,Q^2)$
\item[iii~]LO initial state splitting of $P_{\gamma e}$ 
at $C^{(1,1)}_{e \gamma}(y,Q^2)$
\item[iv~]LO final   state splitting of $P_{e \gamma}$ at 
$C^{(1,1)}_{\gamma e}(y,Q^2)$
\item[v~]NLO initial and  final state radiation off 
$C^{(0,0)}_{ee}(y,Q^2)$
\end{enumerate}
}

\vspace{1mm}
\noindent
The function $C^{(2,1)}(y,Q^2)$ is given by
\begin{eqnarray}
\label{C2SUM}
C^{(2,1)}(y,Q^2) = \sum_{i = {\sf i}}^{\sf v} C_i^{(2,1)}(z,y_h)~.
\end{eqnarray}
The contribution $C^{(2,1)}_{\sf i}(y,Q^2)$ is 
\begin{eqnarray}
\label{C21i}
&&\hspace{-0.7cm}
C^{(2,1)}_{\sf i}(y,Q^2)
\\&&
=\int_0^1 dz P_{ee}^0\left[\theta(z-z_0^I)
J^I \widetilde{C}^{(1,1)}_I-{C}^{(1,1)}\right] \nonumber\\
&&+
\int_0^1 dz P_{ee}^0\left[\theta(z-z_0^F)
J^F\widetilde{C}^{(1,1)}_F-{C}^{(1,1)}\right],\non
\end{eqnarray}
where $C^{(1,1)}(y,Q^2)$ denotes the non--logarithmic part of the
$O(\alpha)$ correction~\cite{ABKR,JBHK2} and $P_{ee}^0(z)$ is the
fermion--fermion LO splitting function
\begin{eqnarray}
\label{P0ee}
P_{ee}^0(z)      = \frac{1+z^2}{1-z}~.
\end{eqnarray}
Also the LO off-diagonal splitting functions
\begin{eqnarray}
\label{P0OD}
P_{e\gamma}^0(z) &=& z^2+(1-z)^2 \\
P_{\gamma e}^0(z)&=& \frac{1+(1-z)^2}{z}
\end{eqnarray}
occur in other contributions to $C^{(2,1)}$. Here both for LO and NLO
splitting functions for equal particle transitions we write the 
contributions for $z < 1$ and account for the +-functions in explicit 
form below.

We express the final result in terms of $\alpha(m_e^2)$ and do therfore
rewrite the coupling constant by
\begin{eqnarray}
\label{RUN}
\alpha(\mu^2) = \alpha(m_e^2) \left[1 -\frac{\beta_0}{4\pi} 
\alpha(m_e^2) \left(\frac{\mu^2}{m_e^2}\right)
\right]~,
\end{eqnarray}
with $\beta_0 = -4/3$. Due to this $C^{(1,1)}$ receives the
running coupling correction
\begin{eqnarray}
\label{C2ii}
C^{(2,1)}_{\sf ii}(y,Q^2) =  - \frac{\beta_0}{2} C^{(1,1)}(y,Q^2)~.
\end{eqnarray}
The contributions $C^{(2,1)}_{\sf iii,iv}(y,Q^2)$ refer to two new
$O(\alpha)$ cross sections~: $d^2 \sigma^{\gamma e,(1)}/dydQ^2$
and $d^2 \sigma^{e \gamma,(1)}/dydQ^2$. The corrections are
\begin{eqnarray}
\label{C2iii}
&&\hspace{-0.7cm}C^{(2,1)}_{\sf iii}(y,Q^2)=\int_{z_0^I}^1 dz 
P_{\gamma e}^0(z)
J^I(z) \widetilde{C}^{(1,1)}_{e\gamma,I}(y,Q^2)\non\\
\\
\label{C2iv}
&&\hspace{-0.7cm}C^{(2,1)}_{\sf iv}(y,Q^2)=\int_{z_0^F}^1 dz P_{e \gamma}^0(z)
J^F(z) \widetilde{C}^{(1,1)}_{\gamma e,F}(y,Q^2)~.\non\\
\end{eqnarray}
The $O(\alpha)$ sub--system cross sections read, see Ref.~\cite{JBHK2}~:
\begin{eqnarray}
&&\hspace{-0.7cm}
\frac{d^2 \sigma^{\gamma e,(1)}}{dydQ^2}=
\frac{\alpha}{2\pi}\sum_{n=0,1}\ln^{1-n}\left(\frac{Q^2}{m_e^2}\right)
C^{(1,n)}_{\gamma e}(y,Q^2)\non\\
\\
&&\hspace{-0.7cm}
\frac{d^2 \sigma^{e \gamma,(1)}}{dydQ^2} =
\frac{\alpha}{2\pi}\sum_{n=0,1}\ln^{1-n}\left(\frac{Q^2}{m_e^2}\right)
C^{(1,n)}_{e \gamma}(y,Q^2)\non\\
\end{eqnarray}
Here, the functions $C^{(1,0)}_{ij}(y,Q^2)$ are given by
\begin{eqnarray}
&&\hspace{-0.7cm}
C_{e\gamma}^{(1,0)}(y,Q^2)= \int_{z_0^I}^1 dz P_{e\gamma}^0(z)J^I(z)
\widetilde{C}^{(0,0)}_I(y,Q^2)\non\\\\
&&\hspace{-0.7cm}
C_{\gamma e}^{(1,0)}(y,Q^2)= \int_{z_0^F}^1 dz P_{\gamma e}^0(z)J^F(z)
\widetilde{C}^{(0,0)}_F(y,Q^2)~.\non\\
\end{eqnarray}
The contribution $C^{(2,1)}_{\sf v}(y,Q^2)$ reads~:
\begin{eqnarray}
\label{eqCv}
&&\hspace{-0.7cm}
C^{(2,1)}_{\sf v}(y,Q^2)\nonumber\\
&&\hspace{-0.2cm}
=  
\int_0^1 
P_{ee,S}^{1,NS,{\rm OM}}\left[\theta\left(z-z_0^I\right)J^I
\widetilde{C}^{(0,0)}_I-C^{(0,0)}\right] \nonumber \\
&&\hspace{-0.3cm}
+\int_{z_0^I}^1
P_{ee,S}^{1,PS,{\rm OM}} J^I \widetilde{C}^{(0,0)}_I \nonumber\\
&&\hspace{-0.3cm}
+\int_0^1
P_{ee,T}^{1,NS,{\rm OM}} \left[\theta\left(z-z_0^F\right) J^F
\widetilde{C}^{(0,0)}_F - C^{(0,0)}\right] \nonumber\\
&&\hspace{-0.3cm}
+\int_{z_0^F}^1
P_{ee,T}^{1,PS,{\rm OM}} J^F \widetilde{C}^{(0,0)}_F
\end{eqnarray}
The splitting functions in the OMS are obtained from the 
$\overline{\rm MS}$--splitting functions \cite{NLOSP} by
\begin{eqnarray}
\label{SPOMS}
&&\hspace{-0.7cm}
P_{ee,S,T}^{1,NS,{\rm OM}}(z)
= P_{ee,S,T}^{1,NS,\overline{\rm MS}}(z) 
+ \frac{\beta_0}{2} \Gamma_{ee}^{0,S,T}(z)~,
\end{eqnarray}
where
\begin{eqnarray}
&&\hspace{-0.7cm}
\Gamma_{ee}^{0,S,T}(z)=-2 \left[\frac{1+z^2}{1-z}\left(\ln(1-z)
+\frac{1}{2}\right)\right]~,
\end{eqnarray}
and 
$P_{ee,S,T}^{1,PS,{\rm OM}}(z) = P_{ee,S,T}^{1,PS,\overline{\rm MS}}(z)$.
The details of the calculation are given in \cite{JBHK2}.
\section{Conclusions}

\noindent
We calculated the $O(\alpha^2 L)$ leptonic QED corrections to
deep inelastic $ep$ scattering for the case of mixed variables. The
corrections are given in terms of double--differential distributions to 
be compared to the double differential Born cross section. 
The calculation
was performed using the renormalization--group decomposition of the
2--loop corrections to the differential cross section w.r.t. mass
factorization in the OMS scheme for the light fermion mass. By this method
an artificial factorization scale $\mu^2$ is introduced on
which the physical cross section does not depend. Its elimination leads
to a re--organization of the cross section which
allows to assemble it in terms of pieces which can be calculated first
individually. We grouped the NLO correction into five terms~: the LO
ISR and FSR radiation correction of the non--logarithmic $O(\alpha)$ 
contribution, a respective term due to charge renormalization, two new 
terms containing $e-\gamma-e$ initial and final state transitions, 
and the ISR and FSR OMS--NLO radiation correction to the Born term.

\end{document}